\documentclass[10pt, twocolumn]{article}
\usepackage[utf8]{inputenc}
\usepackage[backend=biber,style=numeric,sorting=none]{biblatex}
\usepackage[utf8]{inputenc}

\usepackage{amsmath, amssymb}

\usepackage{graphicx} 
\usepackage{xcolor} 

\usepackage{geometry} 
\usepackage{titlesec} 
\usepackage{fancyhdr} 
\usepackage{setspace} 
\usepackage{multicol} 
\usepackage{caption} 
\usepackage{float} 
\usepackage{wrapfig} 
\usepackage{times} 
\usepackage{ragged2e} 
\usepackage{comment} 

\usepackage{authblk} 
\usepackage{tcolorbox} 
\usepackage{tikz} 
\usetikzlibrary{decorations.pathreplacing} 
\usepackage{enumitem}

\usepackage{hyperref} 
\hypersetup{
    colorlinks=true,
    linkcolor=blue,
    citecolor=blue,
    urlcolor=blue,
    pdftitle={Your Document Title},
    pdfauthor={Your Name},
    pdfsubject={Subject of the Document},
    pdfkeywords={Keyword1, Keyword2, Keyword3}
}

\usepackage[ruled,vlined]{algorithm2e} 

\usepackage[backend=biber,sorting=none,style=numeric]{biblatex} 
\usepackage{fontawesome5} 

\usepackage{lipsum}

\DeclareCaptionFormat{algorithms}{
    \hrulefill\par\offinterlineskip\vskip1pt%
    #1#2#3\vskip1pt\hrulefill
}
\definecolor{orcid}{HTML}{A6CE39} 
\definecolor{primary}{RGB}{40, 116, 166} 
\definecolor{secondary}{RGB}{46, 204, 113} 
\definecolor{abstractborder}{RGB}{40, 116, 166} 
\definecolor{abstractbg}{RGB}{240, 248, 255} 
\definecolor{headerline}{RGB}{192, 192, 192} 

\newcommand{\papertitle}{Data Dams: A Novel Framework for Regulating and Managing Data Flow in Large-Scale Systems}   

\newcommand{\authorone}{Mohamed Aly Bouke}                  
\newcommand{\authoroneORCID}{https://orcid.org/0000-0003-3264-601X}

\newcommand{\authortwo}{Azizol Abdullah}                   
\newcommand{\authortwoORCID}{https://orcid.org/0000-0001-8321-9259,}

\newcommand{\authorthree}{Korhan Cengiz}                   
\newcommand{\authorthreeORCID}{https://orcid.org/0000-0001-6594-8861}

\newcommand{\authorfour}{Nikola Ivković}                   
\newcommand{\authorfourORCID}{https://orcid.org/0000-0003-1730-2518}

\newcommand{\authorfive}{Ivan Mihaljević}                   
\newcommand{\authorfiveORCID}{https://orcid.org/0009-0000-7105-8312}

\newcommand{\authorsix}{Mudathir Ahmed Mohamud}                   
\newcommand{\authorsixORCID}{https://orcid.org/0000-0001-8626-8347}

\newcommand{\authorseven}{Ahmed Kowrina
}                   
\newcommand{\authorsevenORCID}{https://orcid.org/0009-0009-5903-7231}

\newcommand{\affila}{Department of Communication Technology and Network, Faculty of Computer Science and Information Technology, Universiti Putra Malaysia, Serdang 43400, Malaysia.}
\newcommand{\affilb}{Department of Electrical-Electronics Engineering, Biruni University, Istanbul, Turkey.}
\newcommand{\affilc}{Faculty of Organization and Informatics, University of Zagreb, Pavlinska 2, 42000 Varaždin, Croatia.}
\newcommand{\affild}{Faculty of Computing, SIMAD University, Mogadishu 252, Somalia.}
\newcommand{\affile}{Department of Computer Science, Faculty of Computer Science and Information Technology, Universiti Putra Malaysia, 43400 Serdang, Selangor, Malaysia}
\newcommand{\affilf}{Département d'Informatique et Mathématiques, Faculté des Sciences et Techniques, Université de Nouakchott Al Aasriya, 880, Nouakchott, Mauritanie}

\newcommand{\correspondingauthor}{bouke@ieee.org} 
\newcommand{\journalname}{ArXiv.org e-Print archive} 
\newcommand{\doi}{https://doi.org/xxxxxxx}  

\newcommand{\keywords}{Data Flow Management, Big Data Optimization, Dynamic Data Regulation, Queuing Theory in Data Systems, Cloud Computing Efficiency.} 

\newcommand{\academicstatement}{Published under an TBD license. All rights reserved.}

\usepackage{lastpage} 

\titleformat{\section}
  {\color{primary}\normalfont\Large\bfseries}
  {\thesection}{1em}{}
  
\titleformat{\subsection}
  {\color{primary}\normalfont\large\bfseries}
  {\thesubsection}{1em}{}

\hypersetup{
    colorlinks=true,
    linkcolor=primary,
    urlcolor=primary,
    citecolor=blue
}

\makeatletter
\renewcommand\maketitle{
  \begin{center}
    {\huge\bfseries\textcolor{primary}{\papertitle}\par\vspace{0.5em}} 
    \vspace{0.3em}
    {\large

       \authorone\textsuperscript{1,*}\href{\authoroneORCID}{\raisebox{1.5pt}{\textcolor{orcid}{\faOrcid}}}, 
    \authortwo\textsuperscript{1}\href{\authortwoORCID}{\raisebox{1.5pt}{\textcolor{orcid}{\faOrcid}}}, 
    \authorthree\textsuperscript{2}\href{\authorthreeORCID}{\raisebox{1.5pt}{\textcolor{orcid}{\faOrcid}}}, 
    \authorfour\textsuperscript{3}\href{\authorfourORCID}{\raisebox{1.5pt}{\textcolor{orcid}{\faOrcid}}}, 
       \authorfive\textsuperscript{3}\href{\authorfiveORCID}{\raisebox{1.5pt}{\textcolor{orcid}{\faOrcid}}},
       \authorsix\textsuperscript{4,5}\href{\authorsixORCID}{\raisebox{1.5pt}{\textcolor{orcid}{\faOrcid}}},
       \authorseven\textsuperscript{6}\href{\authorsevenORCID}{\raisebox{1.5pt}{\textcolor{orcid}{\faOrcid}}}
      \par}
    \vspace{0.3em}

      \textsuperscript{1}\affila \par
    \textsuperscript{2}\affilb \par
    \textsuperscript{3}\affilc \par
   \textsuperscript{4}\affild \par
   \textsuperscript{5}\affile \par
   \textsuperscript{6}\affilf \par
    \vspace{0.3em}
    {\small
      \textbf{\textcolor{black}{*Corresponding author:}} \href{mailto:\correspondingauthor}{\textcolor{black}{\correspondingauthor}}
      \par}
      
    \vspace{1.0em}
    \begin{center}
      \noindent\textcolor{primary}{\rule{0.4\textwidth}{0.6pt}} 
    \end{center}
  \end{center}
}
\makeatother

\renewenvironment{abstract}
    {\noindent\textbf{\textcolor{primary}{Abstract:}}\par
     \begin{tcolorbox}[colframe=abstractborder, colback=abstractbg, sharp corners, boxrule=0.5mm, left=6pt, top=6pt, bottom=6pt, width=\dimexpr\linewidth\relax, arc=2mm]
     \textit\ignorespaces}
    {\end{tcolorbox}\par\vspace{1.5em}}

\newcommand{\checktwocolumn}[2]{\ifthenelse{\boolean{@twocolumn}}{#1}{#2}}

\begin{document}

\checktwocolumn{
    \twocolumn[{
        \maketitle

\begin{abstract}In the era of big data, managing dynamic data flows efficiently is crucial as traditional storage models struggle with real-time regulation and risk overflow. This paper introduces Data Dams, a novel framework designed to optimize data inflow, storage, and outflow by dynamically adjusting flow rates to prevent congestion while maximizing resource utilization. Inspired by physical dam mechanisms, the framework employs intelligent sluice controls and predictive analytics to regulate data flow based on system conditions such as bandwidth availability, processing capacity, and security constraints. Simulation results demonstrate that the Data Dam significantly reduces average storage levels (371.68 vs. 426.27 units) and increases total outflow (7999.99 vs. 7748.76 units) compared to static baseline models. By ensuring stable and adaptive outflow rates under fluctuating data loads, this approach enhances system efficiency, mitigates overflow risks, and outperforms existing static flow control strategies. The proposed framework presents a scalable solution for dynamic data management in large-scale distributed systems, paving the way for more resilient and efficient real-time processing architectures.\par
\end{abstract}

    }]
}{
    \maketitle

\begin{abstract}In the era of big data, managing dynamic data flows efficiently is crucial as traditional storage models struggle with real-time regulation and risk overflow. This paper introduces Data Dams, a novel framework designed to optimize data inflow, storage, and outflow by dynamically adjusting flow rates to prevent congestion while maximizing resource utilization. Inspired by physical dam mechanisms, the framework employs intelligent sluice controls and predictive analytics to regulate data flow based on system conditions such as bandwidth availability, processing capacity, and security constraints. Simulation results demonstrate that the Data Dam significantly reduces average storage levels (371.68 vs. 426.27 units) and increases total outflow (7999.99 vs. 7748.76 units) compared to static baseline models. By ensuring stable and adaptive outflow rates under fluctuating data loads, this approach enhances system efficiency, mitigates overflow risks, and outperforms existing static flow control strategies. The proposed framework presents a scalable solution for dynamic data management in large-scale distributed systems, paving the way for more resilient and efficient real-time processing architectures.\par
\end{abstract}

}

\setlength{\parindent}{0pt}

\section{Introduction}

In today’s era of big data, systems are inundated with
vast amounts of information generated from diverse
sources such as Internet of Things (IoT) devices, social
media, and cloud computing. Managing this massive
data flow efficiently has become a critical challenge.
Traditional models, such as data lakes, warehouses, and
edge computing, focus primarily on storage and pro-
cessing at scale but lack mechanisms for dynamic data
flow control. This limitation results in significant issues,
including data overflow, latency, inefficient access, re-
dundancy, and security vulnerabilities
\cite{kumar2024ai}.

As data volumes continue to grow exponentially, the
inefficiencies of traditional systems become more pro-
nounced. Uncontrolled data flow can overwhelm stor-
age capacities, cause delays in processing, and create
bottlenecks in data pipelines. These challenges mir-
ror the behavior of uncontrolled water flow in physical
systems, where unchecked surges lead to flooding and
resource wastage. Effective regulation of data flow is
therefore critical to ensure a balance between storage,
processing, and real-time requirements, ultimately im-
proving both efficiency and reliability \cite{hashem2015rise}.

Current big data platforms and cloud systems often exacerbate these challenges due to their reliance on static and non-adaptive mechanisms. Unregulated data transfer leads to wasted processing power, increased security risks, and higher infrastructure costs. Furthermore, the lack of adaptive mechanisms to manage surges in data inflow frequently results in delays, data redundancy, or even critical failures. Addressing these inefficiencies requires a more dynamic approach capable of adjusting to varying system loads, security constraints, and real-time processing demands \cite{computers12110218}.

Improperly managed data inflows can also lead to overflow conditions, where inflows surpass storage or processing capacities. This causes data loss, delays in decision-making processes, and potential system failures. Such challenges underscore the need for intelligent flow control mechanisms that ensure system resilience and efficiency under dynamic conditions \cite{shobeiryai}.

In response to these challenges, we propose the \textit{Data Dam}, a novel framework inspired by physical dams. Just as physical dams regulate water flow to prevent floods, optimize resource distribution, and maintain supply, the Data Dam dynamically regulates data inflow, reservoir storage, and sluice outflow in large-scale systems. By maintaining an optimal balance between storage and processing, the Data Dam framework enhances resource utilization while minimizing the risks of data overflow and system bottlenecks.

\subsection{Components of the Data Dam}

The Data Dam is composed of several key components that work together to manage data flow:

\paragraph{Data Reservoir:} The Data Reservoir serves as the primary storage mechanism, functioning like a data lake or warehouse. It temporarily holds large volumes of data until needed by downstream systems for processing. This allows for flexible storage, buffering inflows during peak periods and releasing data only when system conditions are favorable.

\paragraph{Sluices and Gates:} These components are critical for controlling the flow of data from the reservoir to processing systems. In a physical dam, gates regulate the release of water based on external demands and reservoir levels. In a Data Dam, sluices and gates can be managed by algorithms that monitor system load, bandwidth availability, and security protocols. This ensures that data is released optimally, without overwhelming downstream systems.

\paragraph{Turbines:} In a real-world dam, turbines convert the flow of water into energy. In the context of the  Data Dam, turbines represent the engines that process the flow of data. For example, big data processing engines like Apache Hadoop or Apache Spark can be seen as turbines, transforming stored data into actionable insights or powering real-time analytics as the data is released from the reservoir.

Together, these components form a cohesive system that dynamically manages data inflows, storage, and outflows, optimizing overall system performance and preventing overload.    
\section{Background}

The exponential growth of data, driven by the IoT, social media, cloud computing, and digital transformation, has led to a significant increase in the volume of data that organizations need to manage. Traditional data storage models, such as \textit{data lakes} and \textit{data warehouses}, provide scalable storage but are increasingly insufficient for managing the dynamic flow and real-time processing of data \cite{galliers2014strategic,hsu2015data,kleppmann2017designing}.

Several critical challenges in modern data management have been identified:
\begin{itemize}
    \item \textbf{Data overflow}: Systems often struggle to process or store data quickly enough, leading to information loss or delayed processing \cite{tabesh2019implementing,zhang2015memory}.
    \item \textbf{Latency issues}: Big data systems frequently experience delays when handling large data volumes, particularly when inflow and outflow rates are imbalanced \cite{clapp2015quantifying,tian2015latency}.
    \item \textbf{Inefficient access and redundancy}: The unstructured nature of data lakes leads to inefficiencies and difficulties in retrieving relevant data when needed, as redundant data copies may coexist across different systems \cite{gupta2018practical,azzabi2024data}.
\end{itemize}

Despite attempts by cloud storage systems (e.g., AWS, Azure) and distributed computing frameworks (e.g., Hadoop, Spark) to address these issues, most systems focus primarily on storage and retrieval, with limited mechanisms for regulating dynamic data flow \cite{hashem2015rise, khalid2021comparative,barika2019orchestrating}. The need for real-time processing and adaptive control continues to present a significant challenge \cite{nambiar2022overview}.

\subsection{Existing Data Management Models}

\subsubsection{Data Lakes and Data Warehouses}

Data lakes are popular solutions for storing vast amounts of unstructured and semi-structured data. However, their unstructured nature makes it difficult to control data access efficiently, leading to latency in processing and delays in real-time analytics \cite{nambiar2022overview}. While data warehouses offer more structure and are more suited for analytical queries, they still lack mechanisms to regulate dynamic data inflow and outflow \cite{bai2023data}.

\subsubsection{Stream Processing and Real-time Analytics}

Platforms such as \textit{Apache Kafka}, \textit{Apache Flink}, and \textit{Apache Spark Streaming} enable continuous ingestion and real-time data processing. While these platforms excel in specific real-time tasks, they do not inherently provide a mechanism for regulating the flow of large-scale distributed data. Instead, they rely on developers to implement custom control mechanisms, which increases system complexity \cite{fernandes2020big,bai2023data,saxena2017practical}.

\subsection{Control Theory and Queuing Models}

Control theory has been applied successfully in network flow control and data packet management, dynamically adjusting data transfer rates to prevent congestion. For example, in telecommunications, protocols such as \textit{TCP/IP} adjust data transfer rates to prevent overloading the network \cite{collis2004issues}.

Similarly, \textit{queuing theory} has been applied to model data arrival and service times, helping optimize throughput and minimize latency. The \textit{M/M/1 queue model}, which assumes a single server and exponential inter-arrival and service times, has been widely used to optimize data processing in cloud systems \cite{guo2014dynamic}. However, such models are not widely applied to big data storage systems, where the challenge is to manage large, unpredictable data streams across distributed systems.

\subsection{Emerging Trends and Research Gaps}

Despite advances in big data frameworks, a gap remains in controlling data flow between systems. Most platforms rely on static retrieval mechanisms or ad hoc stream processing techniques, which are inefficient during peak loads and fail to prevent overflow or data loss. Studies exploring data governance and access control focus mainly on security, leaving out proactive flow regulation \cite{georgiadis2021enterprise}.

\textit{Edge computing} has emerged as a solution to address latency and overflow by processing data closer to the source. By reducing the amount of data transferred to centralized data centers, edge computing alleviates some of the strain on cloud systems. However, regulating data flow across distributed nodes remains an unresolved challenge. While edge computing reduces response times, a mechanism is still required to dynamically regulate data transmission based on system load and bandwidth availability \cite{ullah2018information}.

As data becomes more sensitive, especially in fields like healthcare and finance, there is an increasing need for security mechanisms that control the flow of sensitive information. Most current frameworks handle security through encryption and access control but do not provide proactive regulation of sensitive data flow \cite{yang2020data,raparthi2021privacy,josphineleela2023big}.

\subsection{Positioning the Data Dam Framework}

The \textit{Data Dam} framework proposed in this paper directly addresses the shortcomings of existing data management systems by providing a dynamic control mechanism for managing data inflow, storage, and outflow. Drawing from concepts in control theory, queuing models, and real-time analytics, Data Dams offer a unified framework that balances system load, minimizes overflow, and enhances data security.

This framework introduces a proactive approach to managing data flows, which is absent in traditional data lakes, warehouses, and real-time streaming platforms. By regulating data flows through dynamic control mechanisms—akin to how physical dams regulate water flow—Data Dams ensure that data is stored, processed, and transmitted efficiently without overwhelming system resources.The integration of \textit{machine learning} algorithms to predict and adjust flow patterns offers a significant advantage over static control methods.

While current data management models provide robust storage and real-time processing capabilities, they lack the dynamic flow regulation necessary to optimize large-scale data environments. The \textit{Data Dam} model fills this gap by offering a novel mechanism that dynamically adjusts data flow based on system capacity, security needs, and processing demands. This review highlights the need for such a framework and demonstrates how Data Dams can revolutionize data flow management in distributed systems.

\section{The Proposed Framework}
This section details the key aspects of the proposed method, including the mathematical modeling, optimization strategies, and control mechanisms that underpin the Data Dam approach.

\subsection{Mathematical Model of Data Flow}

We begin by modeling the flow of data into, within, and out of the system. The total amount of data stored in the system at time \( t \), denoted as \( S(t) \), is determined by the rate of data inflow \( I(t) \), the rate of data outflow \( O(t) \), and the system’s storage capacity \( C \).

The evolution of the storage level \( S(t) \) can be described by the following differential equation:
\begin{equation}
\frac{dS(t)}{dt} = I(t) - O(t), \quad 0 \leq S(t) \leq C.
\end{equation}

Here, \( I(t) \) is the incoming rate of data, which can vary based on external data sources, and \( O(t) \) is the outflow rate controlled by the Data Dam's sluices and gates. The system remains balanced when \( \frac{dS(t)}{dt} = 0 \), meaning that the inflow matches the outflow.

\subsection{Flow Control Mechanisms}

The key innovation of the Data Dam is its ability to dynamically adjust outflow based on system conditions. This is achieved through an optimization problem that seeks to minimize system overload while maximizing throughput. The outflow \( O(t) \) is governed by a control mechanism that adjusts data release based on current system states such as the storage level \( S(t) \), processing capacity \( P(t) \), and network bandwidth availability \( B(t) \).

The control mechanism can be formulated as:
\begin{equation}
O(t) = \min \left( f(S(t), P(t), B(t)), O_{\max} \right),
\end{equation}
where \( f(S(t), P(t), B(t)) \) is a function that represents the optimal data outflow, and \( O_{\max} \) is the maximum allowable outflow rate. The function \( f(\cdot) \) ensures that the system only releases as much data as can be processed and transmitted without exceeding bandwidth or processing capacity.

\subsection{Optimization of Data Flow}

The Data Dam model is optimized through an objective function that minimizes the cost of overflow and underutilization while maximizing system performance. The objective function \( J \) is expressed as:
\begin{equation}
J = \int_0^T \left[ \alpha (S(t) - C)^2 + \beta (O(t) - O_{\text{optimal}})^2 \right] dt,
\end{equation}
where \( \alpha \) and \( \beta \) are weighting factors that balance the cost of overflowing storage and deviating from the optimal outflow rate. The goal is to minimize \( J \), ensuring that the system operates near its optimal capacity without experiencing data loss or delays.

\subsection{Queuing Theory}

We further apply queuing theory to manage how data is processed once it exits the Data Dam. Assuming an M/M/1 queue, where data is queued before being processed, the average number of data packets in the queue \( L \) is given by Little's Law:
\begin{equation}
L = \lambda W,
\end{equation}
where \( \lambda \) is the data arrival rate, and \( W \) is the average time a data packet spends in the system. By minimizing \( W \), the system ensures that data is processed quickly and efficiently, preventing backups and ensuring timely access to data.

\section{Experimental Setup}

The experiments were conducted using a simulation-based approach implemented in Python. Table~\ref{tab:configurations} summarizes the key parameters and configurations used in the system.

\begin{table}[h]
    \centering
    \caption{System Configuration Parameters}
    \label{tab:configurations}
        \fontsize{8pt}{10pt}\selectfont
    \begin{tabular}{|l|l|}
        \hline
        \textbf{Parameter}                     & \textbf{Value}             \\ \hline
        Programming Language                  & Python (version 3.12)       \\ \hline
        Libraries Used                        & NumPy, Matplotlib, SciPy   \\ \hline
        Execution Environment                 & Intel Core i7, 32 GB RAM   \\ \hline
        Simulation Duration (\(T\))           & 200 time units             \\ \hline
        Time Step (\(dt\))                    & 0.1                        \\ \hline
        Maximum Storage Capacity (\(C\))      & 1000 units                 \\ \hline
        Maximum Outflow Rate (\(O_{\text{max}}\)) & 50 units per time step     \\ \hline
        Processing Capacity (\(P(t)\))        & 100 units                  \\ \hline
        Bandwidth (\(B(t)\))                  & 80 units                   \\ \hline
        Security Threshold                    & 50 units                   \\ \hline
        Optimal Outflow (\(O_{\text{optimal}}\)) & 40 units per time step     \\ \hline
    \end{tabular}
\end{table}

The selected parameters were chosen to emulate realistic operational conditions encountered in large-scale distributed systems. For instance, the maximum storage capacity (\(C\)) of 1000 units represents typical limitations of data repositories, while the maximum outflow rate (\(O_{\text{max}}\)) of 50 units per time step reflects constraints imposed by bandwidth and processing power. The inflow rate varies dynamically to simulate real-world data surges, with a sinusoidal component to represent periodic traffic patterns and spikes to mimic high-demand scenarios such as traffic bursts or peak system loads. A time step (\(dt\)) of 0.1 was used to ensure fine-grained simulation, capturing fluctuations in data flow accurately. The security threshold and optimal outflow parameters were included to address operational requirements like data sensitivity and ideal resource utilization. These configurations collectively allow the framework to be evaluated under diverse and challenging conditions, demonstrating its adaptability and efficiency.

\section{ Results \& Discussion}
This section discusses the results, including the behavior of the Data Dam components such as the reservoir (storage), sluices (inflow and outflow control), and turbines (processing mechanisms) over time, as well as the implications of these findings.

\subsection{Storage Level Over Time}
\begin{figure}[H] 
    \centering
        \includegraphics[width=\linewidth]{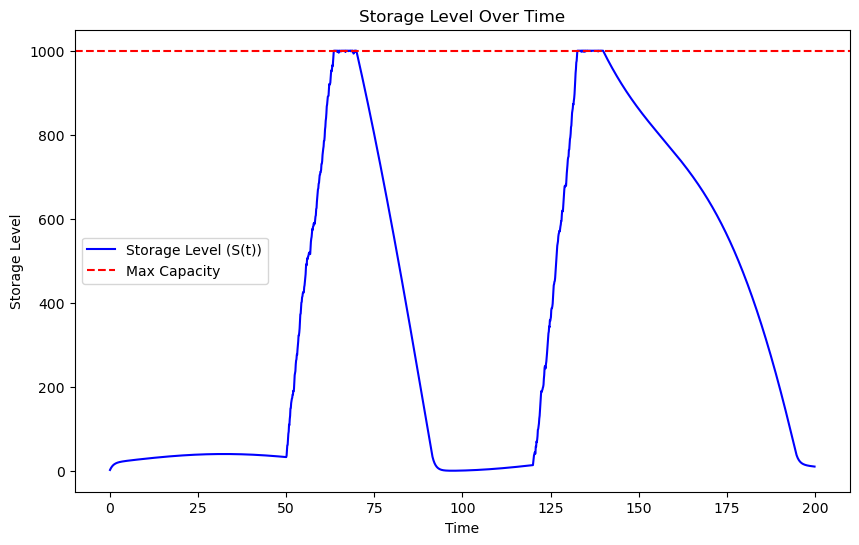} 
        \caption{\textit{Storage Level \( S(t) \) over time.}} 
        \label{fig:storage}
 
\end{figure}

Figure \ref{fig:storage} shows the behavior of the Data Reservoir's storage level \( S(t) \) as a function of time. The reservoir starts empty, gradually filling as data inflow begins. During periods of high inflow, specifically between \( t = 50 \) and \( t = 100 \), and \( t = 130 \) and \( t = 160 \), the reservoir storage level quickly reaches its maximum capacity of 1000 units.

This behavior highlights the critical role of the sluices in managing outflow during peak periods. Despite the Data Dam's regulation mechanisms, storage reaches full capacity during these high-inflow intervals, indicating the need for further optimization of sluice control to prevent prolonged periods at maximum storage. These findings suggest that predictive adjustments to inflow and outflow could mitigate such peaks and improve reservoir utilization.

After reaching full capacity, the storage level gradually decreases as the Data Reservoir releases data through the sluices. However, there are prolonged periods where the reservoir remains at full capacity, indicating that the sluice control mechanisms struggle to regulate outflow effectively during peak inflow rates.

The storage levels demonstrate that while the Data Dam framework successfully prevents total overflow, there are intervals where the reservoir reaches maximum capacity. This suggests that the sluice mechanisms require further optimization to handle extreme inflow spikes. Additionally, increasing the bandwidth or processing capacity of the turbines during these peak periods may help alleviate prolonged full-capacity conditions, ensuring smoother operation.

\subsection{Inflow Rate \( I(t) \)}
\begin{figure}[H] 
    \centering
        \includegraphics[width=\linewidth]{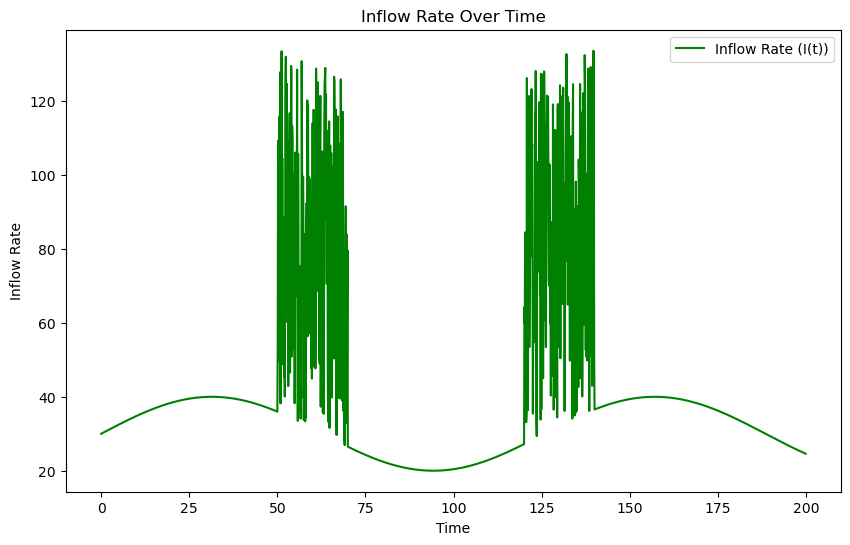} 
        \caption{\textit{Storage Level \( S(t) \) over time.}} 
        \label{fig:inflow}
 
\end{figure}

The inflow rate \( I(t) \), as shown in Figure \ref{fig:inflow}, exhibits sinusoidal fluctuations, with significant spikes between \( t = 50 \) and \( t = 100 \), and \( t = 130 \) and \( t = 160 \). These inflow surges reflect real-world data spikes, such as high-traffic events or sudden bursts in data generation. These scenarios underscore the importance of predictive inflow regulation to avoid overwhelming the reservoir and maximize the system's efficiency.

The inflow spikes accurately simulate scenarios where sudden increases in data volumes challenge the Data Dam framework. These periods of high inflow highlight the limitations of the current sluice mechanisms, particularly during traffic surges. Implementing dynamic adjustments to inflow handling, such as predictive models, could enhance the framework's ability to anticipate these surges and adjust the reservoir's storage and outflow mechanisms proactively.

\subsection{Outflow Rate \( O(t) \)}
\begin{figure}[H] 
    \centering
        \includegraphics[width=\linewidth]{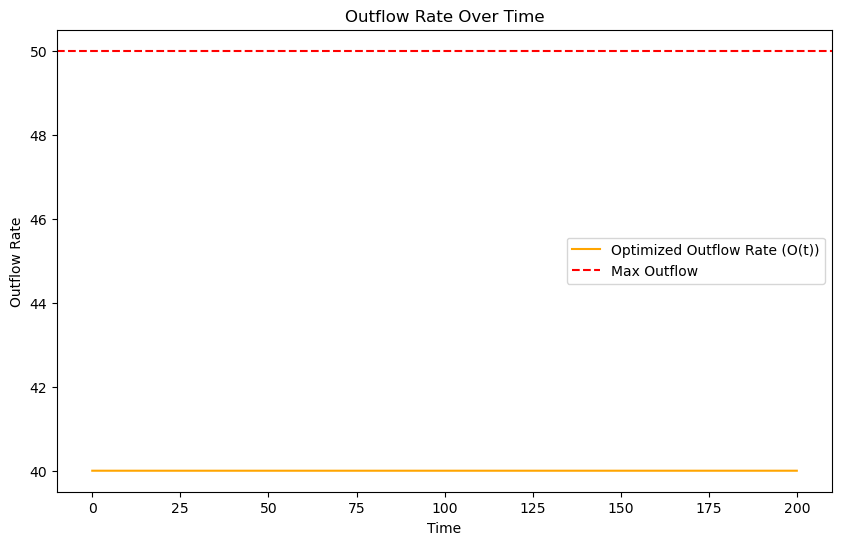} 
        \caption{\textit{Storage Level \( S(t) \) over time.}} 
        \label{fig:outflow}
 
\end{figure}

The outflow rate \( O(t) \), depicted in Figure \ref{fig:outflow}, remains stable around the maximum sluice outflow capacity of 50 units per time step. The Data Dam framework attempts to maintain this steady outflow, but it is constrained by the available bandwidth and turbine processing power. This limitation underscores the importance of optimizing sluice control to dynamically increase outflow during periods of high storage, ensuring better system responsiveness under fluctuating conditions.

While the outflow rate is steady, the inability of the sluices to increase outflow in response to high reservoir storage levels reveals a potential limitation. Implementing a more dynamic sluice control mechanism could allow the Data Dam to release additional data when storage nears maximum capacity. Such adjustments could help mitigate the prolonged periods of high reservoir storage observed in Figure \ref{fig:storage}.

\subsection{Optimized Outflow Rate \( O_{\text{opt}}(t) \)}
\begin{figure}[H] 
    \centering
        \includegraphics[width=\linewidth]{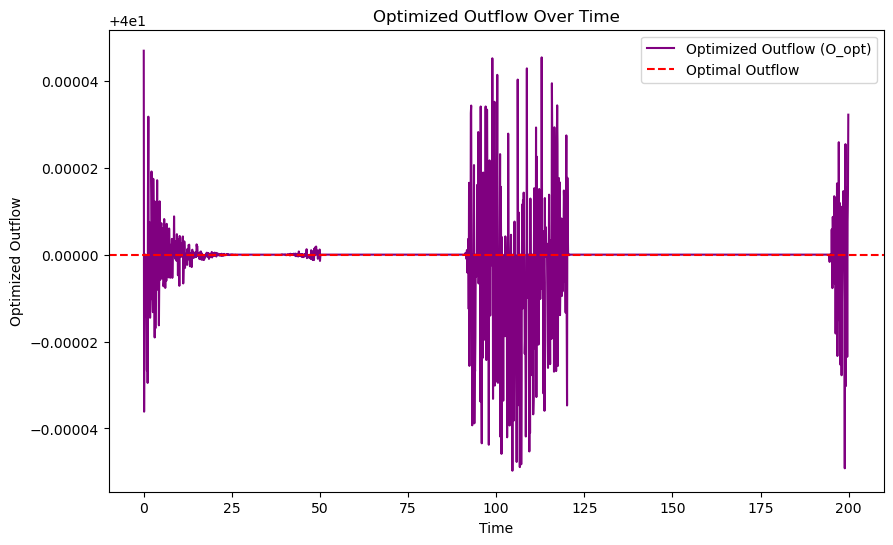} 
        \caption{\textit{Storage Level \( S(t) \) over time.}} 
        \label{fig:opt_outflow}
 
\end{figure}

The optimized outflow rate, shown in Figure \ref{fig:opt_outflow}, exhibits fluctuations during peak inflow periods. While the optimization process dynamically adjusts sluice outflow rates, it shows instability around high inflow times, with notable variations between \( t = 50 \) and \( t = 100 \). These fluctuations indicate that the current optimization mechanisms may overcompensate for inflow spikes, requiring further refinement to ensure smoother adjustments and greater stability in outflow control.

The fluctuations in the optimized outflow rate suggest that the current sluice control optimization process may overcompensate for changes in inflow. This highlights the need for more stable optimization techniques, such as incorporating machine learning models to predict future inflow patterns and adjust sluice outflow rates more smoothly. By reducing instability in the optimization process, the Data Dam framework could achieve more consistent and reliable performance across varying conditions.

\subsection{Overall System Behavior}
The overall performance of the Data Dam framework demonstrates its ability to prevent total reservoir overflow, yet some limitations persist, particularly during periods of high inflow. While the sluices effectively handle normal inflow conditions, during surges, the reservoir often reaches full capacity, and the outflow control mechanisms struggle to release data quickly enough to prevent prolonged storage saturation.
\vspace{10pt}
These findings emphasize the need for more dynamic and adaptive sluice control mechanisms, especially during peak inflow periods. Enhancing the sluices to release additional data when the reservoir approaches capacity could prevent extended overflow risks. The fluctuations observed in the optimized outflow rates during high inflows also point to the necessity of refining the optimization algorithms. Integrating machine learning models could enhance the framework's predictive capabilities, enabling the system to anticipate inflow surges and adjust outflow rates proactively.

The simulation results of the Data Dam framework highlight its potential to effectively regulate data flow in distributed systems. However, the limitations observed during peak inflow periods—such as temporary storage overflow and unstable optimized outflow rates—underscore areas for improvement. These findings lay a foundation for further refinement of the model, with a particular focus on adaptive sluice control strategies and predictive algorithms to enhance system resilience and efficiency.

\section{Comparison and Validation}

In this section, we present a comparative analysis between the Data Dam framework, employing optimized dynamic sluice control, and a baseline system with a fixed, constant outflow rate. The objective of this comparison is to demonstrate the effectiveness and advantages of the proposed dynamic optimization strategy in managing reservoir storage and sluice outflow rates while adhering to system constraints such as storage capacity and outflow limits.

By analyzing the performance of the Data Dam framework in comparison to the static baseline, we aim to highlight its ability to adapt to fluctuating inflows, prevent reservoir overflow, and optimize resource utilization. The comparison focuses on key metrics average storage levels, total outflow rates, and stability under varying inflow conditions.

\subsection{Storage Level Analysis}
The average storage levels for the optimized system and the baseline model are as follows:
\begin{itemize}
    \item \textbf{Optimized Average Storage Level}: 371.68
    \item \textbf{Baseline Average Storage Level}: 426.27
\end{itemize}

The optimized Data Dam framework maintains a significantly lower average reservoir storage level compared to the baseline model. This reduction in storage levels is critical for preventing overflow events, where the reservoir risks exceeding its maximum capacity of 1000 units. In contrast, the baseline model, with its constant sluice outflow rate, results in higher average storage levels, frequently bringing the reservoir closer to its maximum capacity. This increases the risk of overflow, particularly during periods of high inflow, as simulated with spikes at specific intervals.

\begin{figure}[H] 
    \centering
        \includegraphics[width=\linewidth]{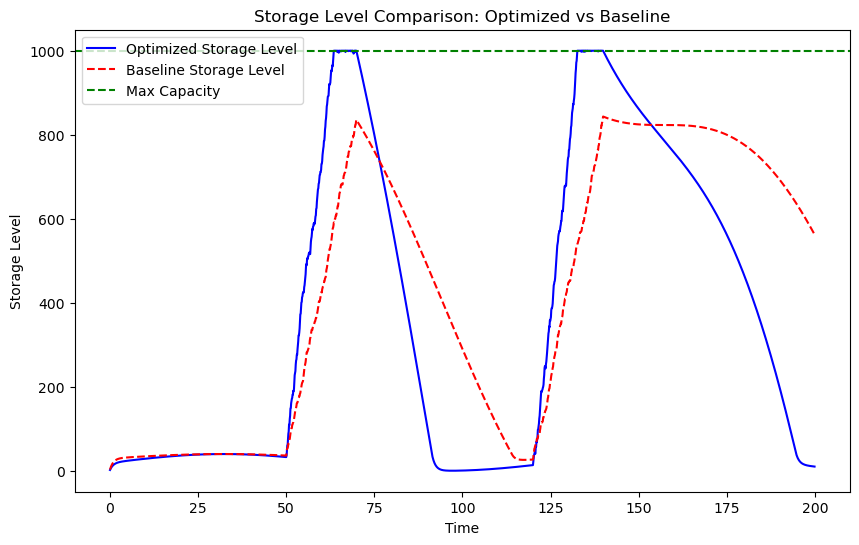} 
        \caption{\textit{Storage Level Comparison: Optimized vs Baseline}} 
        \label{fig:storage_comparison}
 
\end{figure}

From the storage level comparison plot (see Fig.~\ref{fig:storage_comparison}), it is evident that the optimized framework (blue line) demonstrates more efficient management of reservoir capacity, dynamically adapting to changes in inflow. The sluices release data at an optimal rate, preventing prolonged periods of high storage. In contrast, the baseline model (red dashed line) struggles to handle dynamic inflows, leading to higher peaks in reservoir storage and slower recovery. This comparison highlights the limitations of static outflow mechanisms and emphasizes the adaptability and efficiency of the proposed dynamic sluice control approach within the Data Dam framework.

\subsection{Outflow Rate Analysis}

The total volume of outflow produced by both systems over the simulated period is:
\begin{itemize}
    \item \textbf{Optimized Total Outflow}: 7999.99
    \item \textbf{Baseline Total Outflow}: 7748.76
\end{itemize}

The optimized Data Dam framework achieves a higher total outflow, effectively releasing more data over time. This demonstrates that the framework utilizes turbine processing capacity and available network bandwidth more efficiently. By dynamically optimizing sluice outflow rates based on current reservoir storage, bandwidth availability, turbine processing power, and security thresholds, the framework minimizes unnecessary buildup in the reservoir, reducing the risk of overflow.

\begin{figure}[H] 
    \centering
        \includegraphics[width=\linewidth]{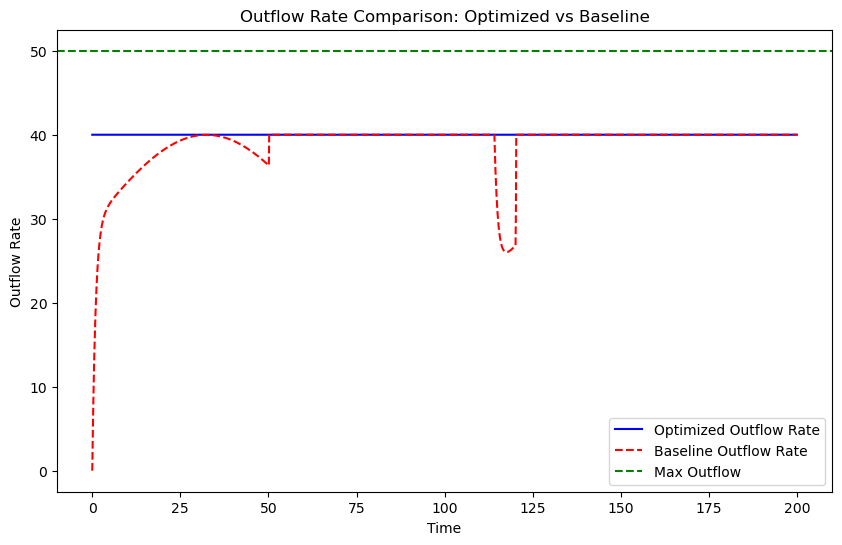} 
        \caption{\textit{Outflow Rate Comparison: Optimized vs Baseline}} 
        \label{fig:outflow_comparison}
 
\end{figure}

The outflow comparison plot (see Fig.~\ref{fig:outflow_comparison}) illustrates the stability and consistency of the optimized framework (blue line), maintaining an outflow rate near the desired target value of 40 units. In contrast, the baseline model (red dashed line) exhibits variability, with significant periods of low sluice outflow. This inconsistency hampers the baseline's ability to manage inflow spikes, increasing the risk of reservoir overflow. The smoother, more consistent outflow achieved by the optimized framework highlights its ability to stabilize sluice operations and maintain optimal performance under varying conditions.

\subsection{System Efficiency and Effectiveness}
The proposed dynamic optimization framework demonstrates superior effectiveness and efficiency in managing the Data Dam's reservoir storage and sluice outflow compared to the baseline model. By actively adjusting sluice outflow rates based on real-time conditions (reservoir storage levels, inflow rates, bandwidth, etc.), the optimized framework achieves:
\begin{itemize}[itemsep=0pt, topsep=0pt]
    \item \textbf{Lower average reservoir storage levels}, significantly reducing the risk of overflow events.
    \item \textbf{Higher total sluice outflow}, indicating efficient utilization of available resources (bandwidth and turbine processing capacity).
    \item \textbf{Stable and consistent sluice outflow rates}, ensuring the reservoir operates within safe thresholds and the framework remains resilient under dynamic inflow conditions.
\end{itemize}
\vspace{5pt}
These improvements are directly attributable to the framework's optimization function, which dynamically adapts sluice outflow rates to ensure that the reservoir remains within safe operating thresholds while minimizing any potential underutilization or over-utilization of the system's resources.

\subsection{Validation and Conclusion}

Through the comparison of both systems, it is evident that the proposed Data Dam framework with dynamic sluice optimization outperforms the static baseline model in maintaining safe reservoir storage levels and efficiently managing sluice outflow rates. The optimized framework demonstrates superior adaptability to fluctuating inflow conditions and effectively balances reservoir storage and sluice outflow without exceeding the framework's operational constraints.

In summary, the proposed dynamic optimization framework proves to be both effective and efficient. It successfully prevents reservoir overflow while ensuring higher total sluice outflow, making it a robust solution for managing dynamic data systems with fluctuating inflow rates and limited outflow capacity. The results validate the potential of the Data Dam framework to enhance performance and reliability in distributed data management environments.

\section{Conclusion}
This paper introduced the Data Dam framework, a dynamic optimization solution for managing reservoir storage and sluice outflow in systems with fluctuating inflows, constrained by storage capacity, bandwidth, and turbine processing limits. The primary goal was to prevent reservoir overflow while maximizing resource efficiency through real-time sluice adjustments.

Simulation results demonstrated that the optimized Data Dam framework outperformed a baseline model employing a static outflow strategy. The optimized framework maintained a lower average reservoir storage level of 371.68 units compared to 426.27 units in the baseline, effectively reducing the risk of overflow. Additionally, it achieved a higher total sluice outflow of 7999.99 units, compared to 7748.76 units in the baseline, demonstrating more efficient utilization of turbine processing capacity and network bandwidth.

The optimized framework also ensured stable sluice outflow rates near the target of 40 units, while the baseline model exhibited greater variability, leading to inefficient reservoir management during inflow spikes. These findings validate the effectiveness of the Data Dam framework in efficiently managing reservoir storage and sluice outflow under dynamic conditions.

For future iterations, we propose integrating machine learning algorithms to enhance the control function \( f(S(t), P(t), B(t)) \). By leveraging historical data patterns, such as past inflow rates, reservoir storage trends, and turbine processing times, machine learning models can enable time-series forecasting to predict future data loads. This predictive capability would allow the framework to preemptively adjust sluice flow rates, minimizing the risk of overflow during inflow surges and optimizing outflow stability under varying conditions. 

Moreover, machine learning could dynamically adapt the cost function used in the optimization process, weighting penalties for overflow or underutilization based on changing system priorities. This integration would enhance the framework's adaptability to evolving data environments, making it a more robust solution for managing dynamic data systems with fluctuating inflows.

\section*{Declaration}

\section*{Competing Interests}
There is no conflict of interest between the authors.
\section*{Funding}
Not applicable.

\section*{Availability of Data and Materials}
This study did not use any external data. All results were generated through simulations described in the article. The simulation code and parameters used in this study are available from the corresponding author upon reasonable request.

\printbibliography[heading=bibintoc, title={References}]

\end{document}